\documentclass[fleqn,usenatbib]{mnras}

\usepackage{newtxtext,newtxmath}
\usepackage{xcolor}

\usepackage[T1]{fontenc}
\usepackage{ae,aecompl}

\usepackage{graphicx}	
\usepackage{amsmath}	

\usepackage{amssymb}	
\usepackage[normalem]{ulem} 

\title[The \textit{[OIII]}$\lambda5007$ EW distribution at z $\sim2$]{
The [OIII]$\lambda5007$ equivalent width distribution at z $\sim2$: The redshift evolution of the extreme emission line galaxies}

\author[K. N. K. Boyett et al.]{Kristan N. K. Boyett,$^{1}$\thanks{E-mail: kristan.boyett@physics.ox.ac.uk (KKB)}
Daniel P. Stark,$^{2}$
Andrew J. Bunker,$^{1}$
Mengtao Tang,$^{3,2}$ 
\newauthor
Michael V. Maseda$^{4}$
\\
$^{1}$Department of Physics, University of Oxford, Denys Wilkinson Building, Keble Road, Oxford OX1 3RH, UK\\
$^{2}$Department of Astronomy/Steward Observatory, University of Arizona, 933 N Cherry Avenue, Tucson, AZ. 85721 USA\\
$^3$Department of Physics and Astronomy, University College London, Gower Street, London WC1E 6BT, UK\\
$^{4}$Department of Astronomy, University of Wisconsin-Madison, 475 N. Charter St., Madison, WI 53706 USA
}

\date{Accepted XXX. Received YYY; in original form ZZZ}

\pubyear{2021}

\begin{document}
\label{firstpage}
\pagerange{\pageref{firstpage}--\pageref{lastpage}}
\maketitle

\begin{abstract}
We determine the [OIII]$\lambda5007$ equivalent width (EW) distribution of $1.700<\rm{z}<2.274$ rest-frame UV-selected (M$_{\rm{UV}}<-19$) star-forming galaxies in the GOODS North and South fields. We make use of deep HDUV broadband photometry catalogues for selection and 3D-HST WFC3/IR grism spectra for measurement of line properties. The [OIII]$\lambda5007$ EW distribution allows us to measure the abundance of extreme emission line galaxies (EELGs) within this population. We model a log-normal distribution to the [OIII]$\lambda5007$ rest-frame equivalent widths of galaxies in our sample, with location parameter $\mu=4.24\pm0.07$ and variance parameter $\sigma= 1.33\pm0.06$. This EW distribution has a mean [OIII]$\lambda5007$ EW of 168$\pm$1\,\AA. The fractions of $\rm{z}\sim2$ rest-UV-selected galaxies with [OIII]$\lambda5007$ EWs greater than $500, 750$ and $1000$\,\AA\ are measured to be $6.8^{+1.0}_{-0.9}\%$,  $3.6^{+0.7}_{-0.6}\%$, and  $2.2^{+0.5}_{-0.4}\%$ respectively. The EELG fractions do not vary strongly with UV luminosity in the range ($-21.6<M_{\rm{UV}}<-19.0$) considered in this paper, consistent with findings at higher redshifts. We compare our results to $\rm{z}\sim5$ and $\rm{z}\sim7$ studies where candidate EELGs have been discovered through {\it Spitzer}/IRAC colours, and we identify rapid evolution with redshift in the fraction of star-forming galaxies observed in an extreme emission line phase (a rise by a factor $\sim10$ between $\rm{z}\sim2$ and  $\rm{z}\sim7$).  This evolution is consistent with an increased incidence of strong bursts in the galaxy population of the reionisation era. While this population makes a sub-dominant contribution of the ionising emissivity at $\rm{z}\simeq2$, EELGs are likely to dominate the ionising output in the reionisation era. 
\end{abstract}




\section{Introduction}
\label{sec:intro}
Over the last decade, considerable  effort has focused on the study of extreme emission line galaxies (EELGs). These systems are often identified through their large [OIII] or H$\beta$ equivalent widths and have been studied in detail at very low redshift \citep{Cardamone09, Amorin10, Izotov11, Brunker20} and in comparable populations at redshifts z $ \sim 1-3$ \citep[e.g.,][]{van_der_Wel_2011, Atek_11, Amorin15, Maseda18, Mengtao19, Du20, Onodera20}. Although rare at the current epoch,  EELGs are thought to represent a significant fraction of the star-forming galaxy (SFG) population at z $>6$ \citep[ e.g.,][]{Smit_2015,deBarros_2019,Endsley20}. They are characterised by a combination of strong nebular line emission and weak rest-optical continuum (i.e. high equivalent width, EW), as expected for a galaxy powered by a very young stellar population with moderately low metallicity.  These systems thus provide a signpost of low mass galaxies as they go through an upturn or burst of star formation.

In the last few years, spectroscopic studies at z $\sim0-2$ have shown that EELGs are more efficient ionising agents than typical star-forming galaxies at these epochs. The production efficiency of ionising radiation $\xi_{\rm{ion}}$ (the measure of the production rate of ionising photons per unit far UV luminosity for a galaxy) increases with [OIII]$\lambda5007$ EWs, reaching its largest values in the most extreme systems within the EELG population (e.g.,  \citealt{Chevallard_2018, Mengtao19, Onodera20}). Many EELGs also show evidence of large ionising photon escape fractions, $f_{\rm{esc}}$ (the fraction of HI ionising photons that reach the inter-galactic medium, IGM, e.g., \citealt{Izotov16,Fletcher19}). Nebular gas under extreme ionisation conditions has been proposed as a necessary but not sufficient criterion for large $f_{\rm{esc}}$ \citep{Nakajima19, Jaskot19, Izotov18}. The ionising conditions are commonly parameterised using the O32 index (the flux ratio of [OIII]$\lambda\lambda$4959,5007 and [OII]$\lambda\lambda$3726,3729), a quantity that will soon be measurable at z $>6$ with {\it JWST}. The O32 values that appear required for large $f_{\rm{esc}}$ (O32$>$6) are uniquely found in EELGs, in particular those systems with [OIII]$\lambda5007$ $\rm{EW}>750$\,\,\AA\, \citep[e.g., ][]{Mengtao19, Sanders20, Onodera20}. At z $\sim3$, \citet{Pahl21} measure an average $f_{\rm{esc}}$ for SFGs to be $\sim 6\%$. In contrast, individual EELGs with [OIII]$\lambda5007$ EW well above 1000\,\AA\, have been observed to exhibit escape fractions up to an order of magnitude greater than these typical systems \citep{Vanzella16,RiveraThorsen17,Izotov18,Fletcher19}. These observations suggest that when galaxies are in an EELG  (or burst) phase, they are likely to contribute more to the ionising background than a typical SFG at z $\simeq 2-3$. However, the fraction of EELGs in the star-forming population is not well constrained at these redshifts. As a result, it is not clear that these systems make a significant contribution to the z $\simeq 2-3$ ionising background.

At higher redshifts (z $>5$), broadband spectral energy distributions (SEDs) suggest that EELGs may be fairly ubiquitous within the SFG population. This inference comes from the presence of strong {\it Spitzer}/IRAC flux excesses in filters that are contaminated by [OIII]+H$\beta$ emission lines. \citep[e.g.,][]{Labb__2013}. The flux excesses imply substantial equivalent widths, often placing galaxies in the EELG regime. While for some of these z $>5$ galaxies the flux excess may also have a  contribution from a Balmer break \citep[e.g.,][]{Eyles05, Eyles07, Roberts-Borsani20}, over some redshift ranges (e.g., $6.6<\rm{z}<6.9$, \citealt{Endsley20, Smit_2014, Smit_2015}) the flux excess can unambiguously be linked to rest-optical emission lines. This has enabled the first glimpse at the [OIII]+H$\beta$ EW distribution in the reionisation era.  At z $>6$, typical [OIII]+H$\beta$ rest-frame EWs are $\sim600-700$\,\AA\, \citep{Endsley20, deBarros_2019, Labb__2013}. \citet{Endsley20} find that 20$\%$ of the SFG population at z $\sim7$ exhibits yet more extreme EWs with [OIII]+H$\beta$ EW $>$ 1200\,\AA\, \citep[see also][]{Smit_2014, Smit_2015, Roberts-Borsani_2016, Castellano17}. Like EELGs at lower redshifts, these large EWs imply the presence of extreme radiation fields, which is further supported by observations of strong line emission from high-ionisation rest-UV metal lines, such as CIII]$\,\lambda1909$\,\AA, CIV$\,\lambda1549$\,\AA\, and HeII$\,\lambda1640$\,\AA\, \citep{Stark17, Stark15a, Stark15b, Mainali17, Laporte_17,Schmidt17, Mainali_18, Hutchison19}.

SFGs are expected to contribute the bulk of the ionising photon budget during the reionisation epoch (e.g., \citealt{Bunker04, Bunker10,Bouwens15b,Robertson15,Stanway16,Stark16, Boylan-Kolchin18,Naidu20,Finkelstein19}; c.f., \citealt{Madau15}). With a large fraction of early galaxies exhibiting extreme [OIII]+H$\beta$ EWs, it is likely that a significant proportion of the ionising output responsible for reionising the IGM comes from SFGs in an EELG or burst phase. Galaxies may also be very effective at producing globular clusters during these intense star formation episodes \citep{Vanzella20,Endsley20}. It is clear that EELGs are likely to play an important role in galaxy growth and reionisation. While existing data hints at the EELG phase becoming more common toward higher redshifts, we currently do not have quantitative constraints on how the prevalence of this population evolves with redshift or varies with galaxy luminosity. This not only hinders our ability to track the contribution of galaxies to reionisation, but it also impedes our understanding of how burstiness may be changing in the galaxy population. 

In this paper, we seek to provide a robust measurement of the [OIII]$\lambda5007$ EW distribution in SFGs at z $\simeq 2$. By selecting our systems in the same manner as those at higher redshifts (i.e., UV-selected dropouts), we aim to provide a baseline measurement which allows the evolution of the EELG population to be established as a function of redshift into the epoch of reionisation. We select galaxies at z $\sim1.7-2.3$ using the HDUV photometric catalogues \citep{Oesch_2018} and we characterise the [OIII]$\lambda5007$ emission line properties using 3D-HST slitless spectra \citep{Momcheva-2016} (described in Section \ref{sec:data}). In Section \ref{method} we determine whether the fraction of the strongest line emitters increases with redshift by comparing our results at z $\sim 2$ with existing measurements at higher redshifts. We discuss implications of our results for reionisation in Section \ref{sec:discussion}. Throughout this paper we assume a $\Lambda$-dominated, flat universe with $\Omega_\Lambda = 0.7$, $\Omega_M = 0.3$, and $H_0=70 $ km s$^{-1}$ Mpc$^{-1}$. All equivalent widths are quoted in the rest-frame and all magnitudes are given in the AB system \citep{Oke83}. 
 
\section{Sample Selection}
\label{sec:data}
\subsection{A photometric sample of $\rm{z}\simeq 1.7-2.3$ galaxies}
Our paper is motivated by recent studies $\rm{z}\sim 7$ that have revealed intense rest-optical nebular emission (e.g., median [OIII]+H$\beta$ EW $\sim600-700$\,\AA, see \citealt{Endsley20, deBarros_2019, Labb__2013}). Whilst examples of such objects have been identified at lower redshifts \citep[z$\sim1-3$][]{Fumagalli12, Atek10, Maseda18}, the fraction of star-forming galaxies with extreme line emission at these redshifts has yet to be quantified. In this paper we combine the HST/WFC3 G141 grism slitless spectroscopy from 3D-HST \citep{Momcheva-2016} with the Hubble Deep UV legacy survey photometry \citep[HDUV;][]{Oesch_2018} to measure the EW distribution of [OIII]$\lambda5007$ of 672 galaxies at redshift $\sim2$ ($1.700 < \rm{z} < 2.274$).  We describe the selection criteria that leads to this sample below.

To facilitate comparison to higher redshift [OIII]+H$\beta$ EW distributions (e.g., \citealt{deBarros_2019,Endsley20}), we select our parent sample from HDUV catalogues \citep{Oesch_2018} using a rest-UV Lyman break dropout colour selection that is similar in nature to those used at z $>4$. The HDUV catalogue adds UV photometry in the F275W and F336W filters over $\sim100$ arcmin$^2$ of the CANDELS GOODS North and South fields. The addition of UV photometry to existing optical and near infrared (NIR) data in these fields allows selection of $\rm{z}\sim2$ galaxies using their characteristic Lyman break \citep{Steidel96, Steidel96b} providing robust photometric redshifts. The two UV filters have average  $5\sigma$ magnitude depths (in 0\farcs4 diameter apertures) of 27.4(27.6) and 27.8(28.0) for F275W and F336W across GOODS North(South) \citep{Oesch_2018}. These two UV filters are combined with the high quality HST ACS optical and WFC3 near-infrared photometry as well as the {\it Spitzer}/IRAC and ground-based filters. \cite{Oesch_2018} use this photometry to generate photometric redshifts for the 30,561 galaxies in the HDUV catalogue using EAZY \citep{Brammer_2008} (using the same procedures as described in \citealt{Skelton_2014}). As we describe below, we construct our sample using these photometric redshifts as our initial selection, similar to that often used to identify $\rm{z}>4$ galaxies (e.g., \citealt{Fink, Mclure11}). At the redshifts we are interested in for this analysis, the photometric redshifts are primarily driven by the Lyman break probed by the HDUV photometry. 

Since our goal is to characterise the strength of the [OIII]$\lambda5007$ emission in these galaxies, we must pick objects that have photometric redshifts which place the [OIII] doublet confidently within the G141 grism spectral window ($\sim$1.0755 to 1.6999 $\mu$m). This translates into a redshift range of $1.148<\rm{z}<2.395$. However there is an additional constraint that limits our selection further. Below z $\sim1.7$, the Lyman limit begins to shift blueward of the F275W and F336W filters, making dropout identification somewhat less reliable without bluer filters (i.e., F225W).  We thus adopt  $\rm{z}=1.700$ as our lower redshift bound. At the high redshift end, we need to choose objects that are confidently within the redshift range ($\rm{z}<2.395$) where we can measure the [OIII] doublet with the G141 grism. Accounting for the typical photometric redshift uncertainty in this sample, we conservatively adopt an upper redshift bound of $\rm{z}=2.274$ for photometric selection, minimising the inclusion of sources with true redshifts above the $\rm{z}=2.395$ threshold. We select all galaxies in the HDUV catalogues with photometric redshifts between $1.700$ and $2.274$. As we will discuss below, sources in this redshift range have SEDs that show strong breaks associated with IGM attenuation, driving the solution of the photometric redshifts in HDUV catalogue. This redshift cut results in a sample of 4026 galaxies.  

We next apply a brightness cut on the rest-UV magnitudes of the photometric sample. This serves two purposes. First, it ensures consistency with higher redshift dropout samples which are traditionally selected in the rest-frame UV.  Second, the magnitude threshold guarantees that our sample is well-matched to the sensitivity of the grism spectra, enabling useful constraints (or upper limits) on the [OIII]$\lambda5007$ EW.   We adopt a fixed cut on  M$_{\rm{UV}}$, the absolute magnitude measured near rest-frame 1500\,\AA. To calculate M$_{\rm{UV}}$ for our sample, we adopt the apparent magnitude in the filter closest to rest-frame 1500\,\AA. We use either the observed F435W (B-band) magnitude below a redshift of $\rm{z}=2.2$ or the observed F606W (V-band) magnitude above $\rm{z}=2.2$. To convert apparent to absolute magnitude, we use the grism-based redshift if the [OIII] doublet is detected at S/N$>$5 and the photometric redshift otherwise. We will show below that the grism redshifts and photometric redshifts are highly consistent. We choose M$_{\rm{UV}}=-19$ as our magnitude cut, ensuring that galaxies are detected at significantly greater than $5\sigma$ in both F435W and F606W filters. This reduces the sample to 766 galaxies. The exact choice of the M$_{\rm{UV}}$ threshold is arbitrary and does not significantly change our best fit parameters for the EW distribution, as discussed further in Section \ref{sec:method_uv_lum}. The M$_{\rm{UV}}$ distribution for our final sample is shown in Figure \ref{fig:Muv_dist}.  

We use the latest grism catalogue (V4.1.5) from 3D-HST \citep{Momcheva-2016, Brammer_2012}. This catalogue contains spectra extracted for all galaxies within the field with a near-infrared (NIR) $JH_{\rm{IR}}$ magnitude (derived from the  J$_{\rm{125}}$ + JH$_{\rm{140}}$ + H$_{\rm{160}}$ combined detection image; see \citealt{Momcheva-2016}) brighter than 26.  There are 28 galaxies within our sample that do not satisfy this NIR threshold and hence do not have available spectroscopic data. To retain these targets within our sample we locate each within the grism slitless spectroscopy and inspect their 2D spectrum, determining that no emission lines were visible in any of the 28 systems. The lack of any visible emission lines is confirmed by utilising an alternative line detection procedure (with no requirements on NIR magnitude) described in \citet{Maseda18}. We will treat each as a non-detection in the following analysis. Keeping these targets within our sample ensures that the NIR magnitude never directly enters our selection, which is important owing to the influence that emission lines can have on the NIR broadband flux if equivalent widths are large. 

We make two final cuts to our photometric sample. First, we remove a small number of sources with very red colours. Such objects are generally excluded from Lyman break selections and would not feature in higher redshift samples we wish to compare to. To remove these objects, we utilise a colour cut of $B-V < 0.8$ (using the F435W and F606W HST ACS filters). This colour threshold is very close to the Lyman break colour cut used at similar redshifts by \citet{Oesch_2018}. This specific colour is chosen to correspond to sources with E(B-V) $<$ 0.4, equivalent to UV slopes with $\beta$ $<$ -1.0 (where $f_\lambda \propto \lambda^{\beta}$). Here we have assumed the dust attenuation law derived from typical $\rm{z}\sim2$ galaxies from \citet{Reddy15}. This B-V threshold reveals 4 very red sources which we remove from our sample. We note that adopting a slightly different colour cut does not significantly alter our results. We also wish to remove sources that may host active galactic nuclei (AGNs), as our goal is to establish the [OIII]$\lambda5007$ EW distribution in star-forming galaxies. We cross-match our sample against deep \textit{Chandra} X-ray imaging across GOODS-N \citep{Alexander_03, Xue_16A} and GOODS-S \citep{Xue_11}, identifying sources that match the coordinates of our sample using a 1\farcs0 search radius. There are 30 targets within our sample that show X-ray counterparts and are removed. Together these two cuts reduce our sample to 732 galaxies.

Our goal is to establish the [OIII]$\lambda$5007 EW distribution in this photometric sample. To do so, we first need to ensure that we have removed any galaxies with grism artefacts from our sample, and second we need to quantify the scatter between the photometric and grism redshifts. We visually examine the grism spectra of our remaining sample to characterise data quality. We identify spectra that have artificial features created by edge effects where the dispersed light of the galaxy falls partially outside the illuminated region of the detector. These artificial features may be mistaken for emission lines, while regions of un-illuminated spectra may contain genuine features that would be missed. The 3D-HST catalogue provides a flag \texttt{\lq f\_cover'}  describing the fraction of dispersed light that falls within the illuminated region of the detector. We find that removing all sources where the \texttt{f\_cover} parameter is below 0.65 effectively eliminates these incidences in our sample. This removes 51 targets. 

We also identify targets that display a negative spectrum on average, where poor contamination subtraction of the contributed light from overlapping spectra has compromised the individual  extractions. The 3D-HST catalogue flag \texttt{\lq f\_negative'} describes the fraction of the spectrum that has a negative flux and we determine that a cut on the \texttt{f\_negative} parameter at 0.7 effectively removes these compromised spectra from our sample, eliminating a further 4 galaxies. We additionally review and remove 1 further system (3D-HST ID S24717) identified to have incorrectly-associated grism emission lines, due to the dispersed light of other galaxies lying coincident with that of the target. We deem the target spectrum to be irretrievable behind the dominant contamination of the neighbour and remove it from the sample. We also individually review all targets with SExtractor \texttt{\lq class\_star'} value greater than 0.5. This inspection results in 3 targets being removed (S20271, S23225, S29694) after being identified as either stars or for having their photometry or slitless spectroscopy compromised by lying in the wings of stellar diffraction spikes. We also remove one object (S24312) as it is associated with one component of a larger galaxy that is already in our sample (S24365). These cuts leave 672 galaxies.

\subsection{Photometric redshifts and contaminants}
\label{sec:contam}
Now that we have a cleaned sample of grism spectra, we can quantify the reliability of the photometric redshifts used to select our parent sample and assess the contamination level among those objects without spectroscopic redshifts. To do so, we take two different approaches. First we identify objects with robust grism redshifts in our photometric sample. We define these as those with confident emission line detections ([OIII] doublet detected at a greater than $5\sigma$ significance) following work in \citet{Mengtao19}. There are 293 galaxies that satisfy this emission line cut in our photometric sample. In this sub-sample, we see that the photometric redshifts do a very good job of reproducing the grism redshifts. To quantify this, we define the typical scatter between photometric and grism redshift using the normalised median absolute deviation, defined as $\sigma_{\rm{NMAD}} = 1.48 \times \rm{median}(\left | \Delta z - \rm{median} (\Delta z) \right | / (1+z_{\rm{spec}}) )$, where $\Delta \rm{z} = z_{\rm{spec}} - z_{\rm{phot}}$ \citep[e.g.,][]{Brammer_2008}. We find a normalised median absolute deviation of $\sigma_{\rm{NMAD}} = 0.037$ for the sub-sample with robust grism redshifts, suggesting good agreement between photometric and grism redshifts. We will assume this dispersion between spectroscopic and photometric redshifts is characteristic of  our total sample, including those systems lacking grism redshifts. This latter subset primarily includes objects with slightly lower equivalent width emission lines. The HDUV photometric redshift 68$\%$ confidence intervals of these sources (median $\Delta $z = 0.063) are comparable to those objects in our sample with grism redshifts (median $\Delta $z = 0.053), likely reflecting their similar continuum magnitudes which in turn enable robust characterisation of the Lyman break (which again is what primarily drives the photometric redshift solution). The average strength of the Lyman break is consistent between the robust grism redshift sub-sample and the sub-sample without significant line detections, both exhibiting a strong F275W-F435W mean break colour of 2.4 mag. Hence in spite of their lower emission line equivalent widths, we expect the photometric redshifts of these sources to be similar in their reliability as those with grism redshifts. 

The typical dispersion between photometric and grism redshifts allows us to calculate the fraction of our sample that is likely to be in-scattered from redshifts where the G141 grism is not able to constrain [OIII] emission ($\rm{z}<1.148$ or $\rm{z}>2.395$). Given the value of $\sigma_{\rm{NMAD}}$, we can perturb the measured photometric redshifts of our sample to quantify this contamination rate. While the in-scatter rate from sources with $\rm{z}<1.148$ is expected to be negligible (owing to the significant buffer provided by our $\rm{z}>1.700$ selection), the dispersion between grism and photometric redshifts suggests that 3.4\% of our photometric sample will have true redshifts of $\rm{z}>2.395$. These sources will not show detections of [OIII] even though they could have strong nebular line emission.  This is easily accounted for in the derived EW distribution, as we will discuss in Section \ref{method}.

In addition to the typical scatter described above, we expect a small number of catastrophic outliers, where the derived photometric redshift is substantially offset from the true redshift of the object. Such objects are generally not included in the Gaussian distribution with $\sigma=\sigma_{\rm{NMAD}}$, so they must be considered separately. Contaminants can be identified in the G141 grism via detection of strong rest-optical lines (H$\alpha$, [OII], or [OIII]) at redshifts well outside of our selected range (1.700$<\rm{z}<$2.274) or through the extensive ground-based spectroscopy that has been conducted in the GOODS fields. We first consider the latter and focus on the GOODS South field where many public redshift surveys have been  conducted (e.g., \citealt{Fevre05,Vanzella08, Balestra10,Kurk13}). We cross-match these surveys with our GOODS-South catalogue (318 galaxies), finding 75 unique matches. The spectra in these catalogues find only one catastrophic outlier with a confident redshift identification (3D-HST ID: 19233, matched to \citealt{Balestra10} ID: J033227.25-274919.2 with $\rm{z}_{\rm{spec}} = 0.5568$), indicating a catastrophic outlier fraction of 1.3\% in the ground-based spectra and suggesting a fraction of at least 0.3\% in our total GOODS-S sample. The WFC3/IR grism spectra provide an independent check. The G141 grism is able to detect H$\alpha$ down to $\rm{z}=0.6$, providing a window on low-z interlopers in our sample. Here we define a catastrophic outlier as a source with a redshift that is further than 5$\sigma_{\rm{NMAD}}$ from our lower redshift bound, corresponding to $\rm{z}=1.2$. We identify 10 sources with H$\alpha$ detections, but none fall below $\rm{z}=1.4$. These redshifts are consistent with the Gaussian distribution implied by our calculation of $\sigma_{\rm{NMAD}}$, hence the grism sample is also suggestive of a low catastrophic outlier rate. In the following section, we will conservatively assume a catastrophic outlier rate of 1.3\% within our sample. We will add this catastrophic contamination fraction to that derived for more typical source-to-source redshift scatter. The total contamination fraction (3.4\%+1.3\%=4.7\%) will be modelled when deriving the EW distribution.

\begin{figure}
    \centering
    \includegraphics[width=\columnwidth]{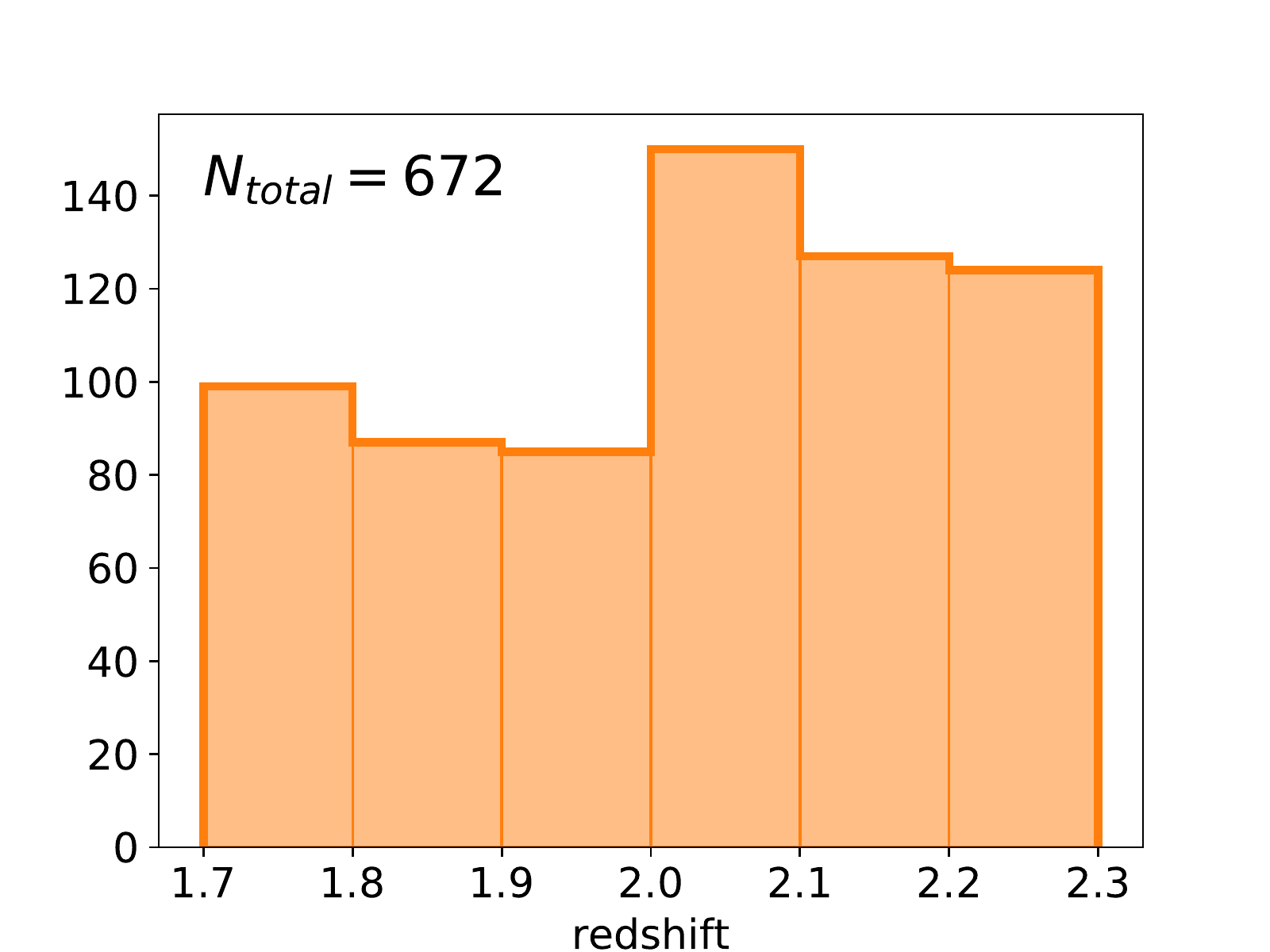}
    \caption{Photometric redshift distribution of the parent sample for our [OIII] EW distribution. The sample includes 672 $\rm{z}\simeq 2$ galaxies in the combined GOODS North and South fields. The systems are selected to have HDUV photometric redshift in the range $1.700<\rm{z}<2.274$ \citep{Oesch_2018} and absolute magnitude brighter than M$_{\rm{UV}}=-19$.}
    \label{fig:red_dist}
\end{figure}

\begin{figure}
    \centering
    \includegraphics[width=\columnwidth]{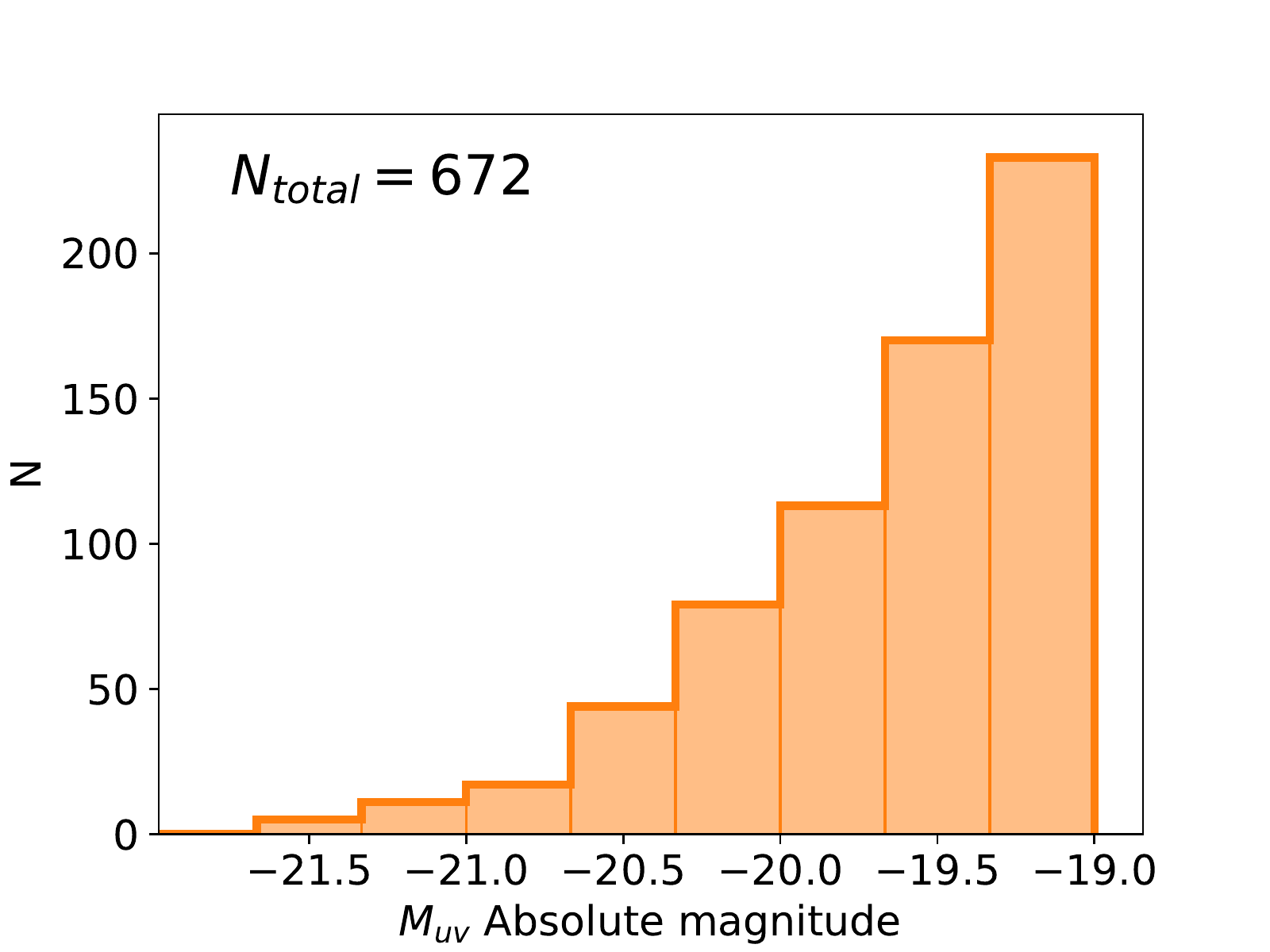}
    \caption{Absolute magnitude distribution of the photometric sample we use for our [OIII] EW distribution. We apply an absolute magnitude threshold of 
    M$_{\rm{UV}}=-19$ to our sample.}
    \label{fig:Muv_dist}
\end{figure}

To summarise, we are left with a final sample of 672 galaxies, with 354 in GOODS North and 318 in GOODS South. The total photometric sample includes 293 systems with $\geq5\sigma$ detections of the [OIII] doublet. The photometric redshift distribution is presented in Figure \ref{fig:red_dist} with a medium value of z=2.0. The absolute M$_{\rm{UV}}$ magnitude distribution is shown in Figure \ref{fig:Muv_dist} and spans $-21.6<$ M$_{\rm{UV}} < -19.0$, with 75.6$\%$ of targets fainter than M$_{\rm{UV}} = -20.0$.  These values are well-matched to those of $\rm{z}\simeq 7$ galaxies that have been used to infer [OIII]+H$\beta$ EWs via {\it Spitzer}/IRAC excesses \citep{deBarros_2019,Endsley20}.

\section{[OIII]$\lambda5007$ Equivalent Width Distribution at $\rm{z}\sim2$} 
\label{method}
In this section, we compute the [OIII]$\lambda5007$ EW distribution at z $\sim2$ from our selected sample. We first discuss measurements of individual EWs in Section \ref{sec:method_EW} and then describe how we characterise the functional form of the EW distribution in Section \ref{sec:method_EW_results}. From this distribution we  measure the fraction of galaxies caught in an extreme EW phase (the EELG fraction), corresponding to very large sSFRs expected when galaxies experience bursts of star formation. We characterise the redshift and luminosity dependence of the EELG fraction in Section \ref{sec:method_red_evo} and Section \ref{sec:method_uv_lum} respectively.

\subsection{Individual equivalent width measurements}
\label{sec:method_EW}
The first step in deriving the [OIII]$\lambda 5007$ EW distribution is to robustly measure the individual EWs for each of the 672 galaxies in our final sample. As described in Section \ref{sec:data}, we create a sub-sample of galaxies with significant [OIII] doublet detections ($\geq5\sigma$) and determine their [OIII]$\lambda 5007$ EW. For the remaining sample we derive upper limits on the EW. We describe this procedure below. 

Our approach to line measurement is similar to that taken in \citet{Mengtao19}. For galaxies with a $\geq5\sigma$ [OIII] doublet detection in the grism spectra of \citet{Momcheva-2016}, we use their measured line flux in determining the EW. We note that the [OIII]$\lambda4959,5007$\,\AA\ doublet is unresolved at the spectral resolution of the G141 grism and the total [OIII] doublet flux is reported in \citet{Momcheva-2016}. We correct the total line flux of the doublet to the expected line flux of [OIII]$\lambda5007$ alone, assuming the theoretical line flux ratio of 2.98 between [OIII]$\lambda5007$\,\AA\ and [OIII]$\lambda4959$\,\,\AA\ \citep{Storey00}. For the estimate of the stellar continuum at the observed wavelength, we derive the continuum flux density from the broadband photometry rather than using the spectral continuum measured from the grism spectra by \citet{Momcheva-2016}. In many objects in 
our sample, the continuum is often very low S/N. This approach allows us 
to have a uniform method for our entire sample, and as we show later in 
this section the two approaches give consistent EW measurements. 
To compute the photometric broadband flux density in the filter covering the [OIII] doublet, we compute the continuum flux using an aperture that is matched to the 
grism spectral extraction aperture for each source. The F125W broadband filter is used to determine the continuum for galaxies below z=1.8, the F140W filter is used in the range z=1.8-2 and the F160W filter is used above z=2 (or above z=1.8 when F140W is unavailable). In determining the continuum flux density from the broadband photometry, we correct the flux within the filter for the contribution of emission lines. We consider the contribution from all lines detected in the grism spectra, weighted by the filter transmission profile at the respective location of each line. For the majority of sources in our sample, this is a very small correction that does not significantly impact the EW inference. We verify that systems with grism non-detections have minimal emission line contribution to the broadband flux ($<$5\%). We find that three sources in our sample are sufficiently line-dominated that the derived continuum flux density (after line subtraction) lies below the 5$\sigma$ flux density limit for that filter. In these cases (S19339, S30532, N16381), we utilise the average of neighbouring broadband filters without line contamination to measure the continuum flux density. Finally we inspect by-eye all galaxies where the \citet {Skelton_2014} half-light radii of the target and a nearby neighbour overlap. Two targets of concern are identified, and we employ Galfit \citep{Peng02} to model and subtract the contribution from the neighbour to the continuum flux density. 

In order to compute the [OIII]$\lambda5007$ EW, we must make a small correction to convert the continuum flux density at the effective wavelength of the chosen filter (i.e., F125W, F140W, F160W) to that at the rest-wavelength of [OIII]$\lambda5007$. Given the small wavelength baseline involved, we assume that the continuum is flat in $f_\nu$, as is appropriate for unreddened stellar populations at the range of ages spanned by our sample \citep[e.g., ][]{Maseda14, van_der_Wel_2011}. This correction changes by less than 1\% if we consider galaxies with more reddening (i.e., selective extinction of E(B-V)=0.2 with a Calzetti reddening curve). The [OIII]$\lambda5007$ EW is then taken as the ratio of the line flux to the continuum flux density and corrected to the rest-frame using the grism-measured spectroscopic redshift (EW$_{\rm{rest}}= $ EW$_{\rm{obs}} / (1+\rm{z})$).

For systems lacking [OIII] doublet detections with S/N$\geq$5, we derive $5\sigma$ upper limits on the [OIII]$\lambda5007$ EW. For each non-detection, we adopt the $5\sigma$ grism flux upper limit using the line sensitivity equation, described in \cite{Momcheva-2016}. This is an empirical parameterisation of the flux uncertainties determined through 2D model fitting of the grism spectra, acting as a function of the filter transmission throughput at the observed wavelength and the aperture size required to span the spatial extent of the galaxy (for each object we adopt the broadband flux radius from \citealt{Skelton_2014}). The EW $5\sigma$ upper limits are then produced by combining the line flux upper limit with the broadband continuum (derived as discussed above) and corrected to the rest-frame using the HDUV photometric redshifts.

Within our photometric sample, the sources with [OIII] doublet detections in the grism spectra have rest-frame [OIII]$\lambda$5007 EWs ranging between 40 and 1800\,\AA\ with a median of 214\,\AA. For sources that are sufficiently bright in the continuum ($JH_{\rm{IR}} < 23$) we are sensitive to low EWs ($<50$\,\AA), allowing us to have a broad [OIII]$\lambda$5007 EW range, but for more typical magnitudes in our sample, the grism spectra are not sufficiently deep to reach such low [OIII]$\lambda$5007 EWs. We find the median of the $5\sigma$ [OIII]$\lambda$5007 EW upper limits to be 93\,\AA. The dynamic range of our sample’s NIR magnitudes means the distribution of EW measurements and upper limits overlap. A histogram of the resultant [OIII]$\lambda5007$ EWs is shown in Figure \ref{fig:Bayes_best_fit}. The detected sources (blue) are assigned to their appropriate EW bin while the contribution to the histogram from each EW upper limit source (orange) is spread over the EW parameter space below the $5\sigma$ upper limit, following the best fit log-normal distribution (discussed in Section \ref{sec:method_EW_results}). The sum of the probability across all bins below the $5\sigma$ upper limit equals 1 for an individual galaxy with no [OIII] detection at $5\sigma$.

\subsection{Equivalent width distribution}
\label{sec:method_EW_results}
As outlined in the introduction, our primary goal is to derive the [OIII]$\lambda5007$ EW distribution. In doing so we seek to quantify what proportion of the z $\sim2$ population is in an extreme EW phase, as is believed to be common at z $>7$. Equipped with robust measurements and upper limits for the EW for our described sample we can now characterise the distribution of EWs. We will model our sample, inclusive of non-detections, with a log-normal EW distribution. This functional form has been shown to be a good fit to other 
rest-optical EW samples \citep[see][]{Lee07, Lee12, Ly11, Schenker_2014, Stark2013, Endsley20}. 

To infer the underlying [OIII]$\lambda5007$ EW distribution from our observed sample we follow the Bayesian method set out by \citet{Schenker_2014} which preserves information on each galaxy's EW measurement uncertainty, while removing the need to bin the equivalent width measurements. The EW distribution is modelled as a log-normal function $\theta = [\mu_{\rm{LN}}, \sigma_{\rm{LN}}]$\footnote{Equivalently this is the same as setting a normal distribution in logarithmic space with parameters $\theta = [\mu_{\rm{N}}, \sigma_{\rm{N}}]$. Both forms have been used in the literature and the relation between the normal distribution parameters and our log-normal parameters is given by; $\sigma_{\rm{N}} = \sigma_{\rm{LN}} \times \log_{10}(e)$ and $\mu_{\rm{N}} = (\mu_{\rm{LN}}-\sigma_{\rm{LN}}^2) \times \log_{10}(e)$ and we will provide the best fit results from our analysis in both formats.}. We follow \citet{Schenker_2014} and place a flat prior $P(\theta)$ over both the log-normal location $\mu_{\rm{LN}}$ and variance $\sigma_{\rm{LN}}$ parameters allowing the posterior to be determined directly as the model likelihood from our observed sample. 

For a set of model parameters [$\mu_{\rm{model}}, \sigma_{\rm{model}}$] the model log-normal probability distribution is given by
\begin{equation}
    P(EW|\theta)_{\rm{model}} = \left ( 2\pi\,\sigma_{\rm{model}}^2 \, EW^2 \right )^{-\frac{1}{2}}
e^{-\frac{(ln(EW) - \mu_{\rm{model}})^2}{2\sigma_{\rm{model}}^2}}
\end{equation}
and the Gaussian measurement uncertainty on the measured equivalent width is given by
\begin{equation}
    P(EW)_{\rm{obs_i}} = \left ( 2\pi\,\sigma_{\rm{obs_i}}^2 \right )^{-\frac{1}{2}} e^{-\frac{(EW - \mu_{\rm{obs_i}})^2}{2\sigma_{\rm{obs_i}}^2}}.
\end{equation}
Where $\mu_{\rm{obs_i}}$ and $\sigma_{\rm{obs_i}}$ are the determined EW and observational uncertainty for the $\rm{i}^{th}$ system. The likelihood over the complete dataset is taken as the product of the individual likelihoods of each galaxy within the sample. The individual likelihood for each detected source ($i$) is defined as
\begin{equation}
    P(obs_{\rm{i}}|\theta)_{\rm{detect}} = \int_{0}^{\infty}P(EW)_{\rm{obs_i}} \cdot P(EW|\theta)_{\rm{model}} \,\, d\, EW.
\end{equation}
The combination of the true EW distribution model with the Gaussian profile describing the EW measurement within the likelihood integral addresses observational noise to avoid overestimating the number of EELGs, where noise would preferentially scatter sources from the bulk of the distribution (with low EWs) towards extreme EW and broaden the observed EW distribution (up-scatter). 

 For sources which are undetected, the individual likelihood is defined by
\begin{equation}
    P(obs_{\rm{i}}|\theta)_{\rm{non\_detect}} = P(EW<EW_{5\sigma_{\rm{i}}}|\theta)
\end{equation}
\begin{equation*}
    \vspace{-0.5cm}
    \;\;\;\;\;\;\;\;\;\;\;\;\;\;\;\;
    \;\;\;\;\;\;\;\;\;\;\;\;\;\;\;\;
    + P(EW>EW_{5\sigma_{\rm{i}}}|\theta) \cdot C_{1, \rm{i}} + C_2
\end{equation*}

where $C_{1, \rm{i}}$ is the proportion of the $\rm{i}^{th}$ spectra that is unilluminated due to its location on the detector. $C_{2}$ is the photometric redshift in-scatter fraction, evalutated in Section \ref{sec:contam} to be 4.7$\%$. This probability is determined as the sum of: the likelihood that the galaxy has a EW below the $5\sigma$ limit; the likelihood that the galaxy has an EW above this threshold multiplied by $C_{1, \rm{i}}$; and $C_2$. Due to the inclusion of the in-scatter term ($C_2$), this probability $P(obs_{\rm{i}}|\theta)_{\rm{non\_detect}}$ may exceed 1 and so to avoid this a maximum value is enforced set equal to 1. 

\begin{table}
\begin{center}
\caption{Best fit model parameters for the rest-frame EW(\,\,\AA). We show the results for the log-normal distribution (LN) and normal distribution in logarithmic space (N).}
\label{tab:mcmc_results}
\begin{tabular}{ccccc}
Number    & $\mu_{\rm{LN}}$        & $\sigma_{\rm{LN}}$ & $\mu_{\rm{N}}$ & $\sigma_{\rm{N}}$     \\ \hline 
\hline
672 & $4.24\pm0.07$ & $1.33\pm0.06$ & $1.08\pm0.10$ & $0.58\pm0.03$
\end{tabular}
\end{center}
\end{table}

\begin{figure*}
    \centering
    \includegraphics[width = 15cm]{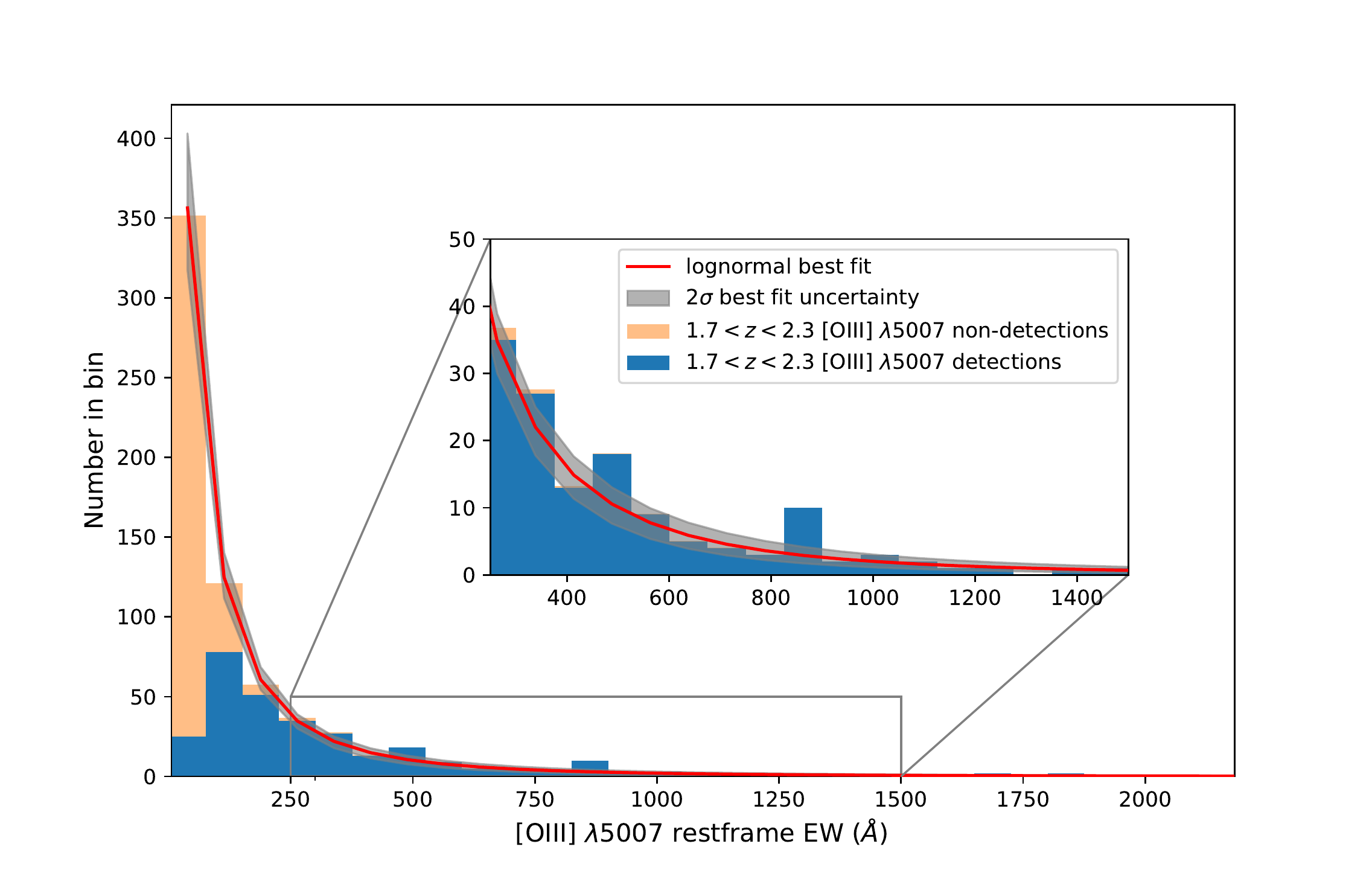}
    \caption{The EW distribution of [OIII]$\lambda5007$ for star-forming galaxies at $1.700 < \rm{z} < 2.274$. The best fit log-normal model (red line) and $2\sigma$ uncertainties (grey) along with the EW histogram in 75\,\AA\ bins for our $M_{\rm{UV}}$ selected sample constructed from $5\sigma$ [OIII] doublet detections (blue) and $5\sigma$ upper limits (orange), where the histogram contribution from each [OIII] doublet non-detection is assigned following the best fit model below the associated $5\sigma$ [OIII]$\lambda5007$ EW upper limit. }
    \label{fig:Bayes_best_fit}
\end{figure*}

A Markov Chain Monte Carlo (MCMC) approach is taken to efficiently cover the parameter space using \texttt{emcee} \citep{Foreman-Mackey13}. The marginalised posterior distributions over the log-normal parameters are used to determine the best fit model parameters, which are presented in Table \ref{tab:mcmc_results}. We over-plot the best fit model for the sample on the EW distribution in Figure \ref{fig:Bayes_best_fit} with the $2\sigma$ model uncertainties indicated by the grey shaded region.

From our best fit results in Table \ref{tab:mcmc_results} we report a mean (with standard error on the mean) and median EW of $168\pm1$\,\AA\ and $70\pm5$\,\AA, with the EELG population skewing the mean EW higher than the median. This mean is within the expected [OIII]$\lambda5007$ EW range (80-250\,\AA) for log$_{10}(\frac{M}{M_\odot}) = 9-10$ stellar mass galaxies based on the z $\sim2$ MOSDEF empirical relation by \citet{Reddy18}. The median is significantly lower than the $\sim 450$\,\AA\, median [OIII]$\lambda5007$ EW inferred for z $>7$ samples \citep{Endsley20, deBarros_2019, Labb__2013} (when converted from [OIII]+H$\beta$ to [OIII]$\lambda5007$ EW assuming the assumptions detailed in Section \ref{sec:method_red_evo}). We note that the median [OIII]$\lambda5007$ EW in the sub-sample where the [OIII] doublet is detected ($\geq5\sigma$) was 214\,\AA\ (see Section \ref{sec:method_EW}), whereas taking the whole sample and treating upper limits as described we measure a lower median [OIII]$\lambda5007$ EW of 70\,\AA. This should be expected since intrinsically low EW sources would be more likely to have [OIII] doublet non-detections. 

\begin{figure*}
    \centering
    \includegraphics[width  = 15cm]{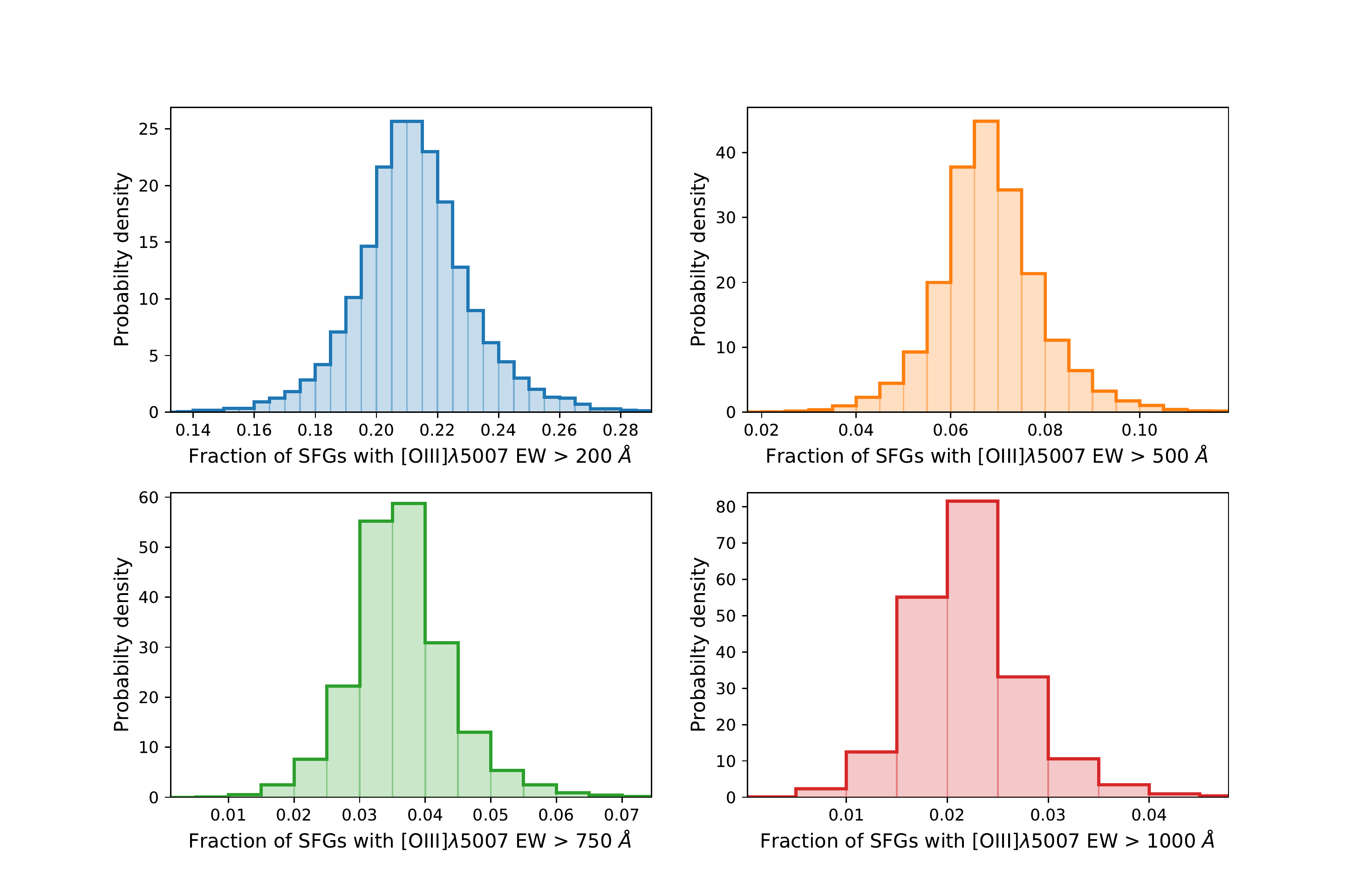}
    \caption{The fraction of star-forming galaxies at z $\sim2$ with extremely large [OIII]$\lambda5007$ EW.  Posterior distributions for four different [OIII]$\lambda5007$ EW thresholds are shown in blue (200\,\AA, top left panel), orange (500\,\AA, top right panel), green (750\,\AA, bottom left panel) and red (1000\,\AA, bottom right panel).}
    \label{fig:EELG_dist}
\end{figure*}

Through fitting a functional form, the proportion of SFGs with an EW above a given threshold can be easily calculated and we shall call this the EELG fraction. For each set of model parameters $\theta$ of the MCMC run the EELG fraction is recorded and the resultant posterior density function (PDF) over all models is then used to estimate the best fit overall EELG fraction and uncertainty for the population (see Figure \ref{fig:EELG_dist}, discussed below). 

Within the literature there is no set definition for the threshold [OIII]$\lambda5007$ EW a galaxy must have to be classified as an EELG and the quoted threshold varies from author to author. Commonly taken threshold values range from [OIII]$\lambda5007$ EW $\sim100-1000$\,\AA. At low redshift, \citet{Amorin15} employ a $\geq100$\,\AA\, threshold to construct an EELG sub-set from a $0.11<\rm{z}<0.93$ SFG sample, and in our distribution this EW threshold results in an EELG fraction of $39^{+2}_{-3}\%$ at z $\sim2$. At intermediate redshifts considerable attention has been focused on photometric selection of EELGs, for example, \citet{van_der_Wel_2011} who identify  z $\sim1.7$ galaxies through HST J-band (F125W) flux excess. These selected EELGs at z $\sim1.7$ tend to have [OIII]$\lambda5007$ EWs above 500\,\AA, which is comparable to the typical [OIII]$\lambda5007$ EW found at z $>7$. Adopting a threshold of 500\,\AA\ yields an EELG fraction of $6.8^{+1.0}_{-0.9}\%$ in our EW distribution at z $\sim2$.

More recently, spectroscopic work at z $\sim2$ exploring the stellar population and gas properties as a function of [OIII]$\lambda5007$ EW  \citep[e.g.][]{Mengtao19, Mengtao20,Du20} has focused on the most extreme line emitters, those with [OIII]$\lambda5007$ EWs above 750 or 1000\,\AA. These are the galaxies at z $\sim2$ found to have the highest $\xi_{\rm{ion}}$, O32 values, and the potential for large ionising escape fractions. In our z $\sim2$ distribution these are rare, accounting for only $3.6^{+0.7}_{-0.6}\%$ above a threshold of 750\,\AA\ and $2.2^{+0.5}_{-0.4}\%$ above a threshold of 1000\,\AA.

In Figure \ref{fig:EELG_dist} we present the EELG fraction posterior density function (PDF) for four [OIII]$\lambda5007$ EW thresholds. These correspond to the rough mean [OIII]$\lambda5007$ EW of SFGs at z $\sim2$ (200\,\AA), the typically EW of SFGs at z $>7$ (500\,\AA) and the EW seen in the most extreme line emitters at z $\sim2$ (750 and 1000\,\AA). The 200\,\AA\, threshold accounts for roughly a fifth of SFGs at z $\sim2$ with the higher EW thresholds recovering diminishing fractions. The 500\,\AA\, threshold highlights that the SFGs common at z $>7$ are rare at z $\sim2$, with the most extreme candidates effectively absent from the z $\sim2$ population. We present the measured EELG fractions for each [OIII]$\lambda5007$ EW threshold in Table \ref{tab:EELG-fraction}.

\begin{table*}
\begin{tabular}{ccccc}
Sample               & EW$\geq200$\,\AA        & EW$\geq500$\,\AA       & EW$\geq750$\,\AA       & EW$\geq1000$\,\AA      \\ \hline
\hline
full sample          & $21.2^{+1.7}_{-1.6}\%$ & $6.8^{+1.0}_{-0.9}\%$ & $3.6^{+0.7}_{-0.6}\%$ & $2.2^{+0.5}_{-0.4}\%$ \\
\\
$1.70 < \rm{z} < 2.01$      & $19.7^{+2.1}_{-2.1}\%$ & $5.4^{+1.2}_{-1.1}\%$ & $2.6^{+0.8}_{-0.9}\%$ & $1.5^{+0.6}_{-0.4}\%$ \\
\\
$2.01 < \rm{z} < 2.274$      & $22.8^{+2.6}_{-2.3}\%$ & $8.2^{+1.6}_{-1.5}\%$ & $4.7^{+1.2}_{-1.1}\%$ & $3.0^{+0.9}_{-0.8}\%$ \\
\\
$-19.5 < M_{\rm{UV}} < -19$ & $20.1^{+2.4}_{-2.1}\%$ & $6.7^{+1.4}_{-1.3}\%$ & $3.6^{+1.0}_{-0.9}\%$ & $2.3^{+0.8}_{-0.7}\%$ \\
\\
$-21.6 < M_{\rm{UV}} < -19.5$ & $22.3^{+2.4}_{-2.3}\%$ & $7.1^{+1.5}_{-1.3}\%$ & $3.7^{+1.0}_{-0.8}\%$ & $2.2^{+0.7}_{-0.6}\%$
\end{tabular}
\caption{The fraction of star-forming galaxies in an extreme emission line phase (the EELG fraction) with [OIII]$\lambda5007$ EW above four rest-frame EW thresholds. The EELG fractions are presented as a percentage ($\%$) for the full z $\sim2$ UV-selected sample, the sample split into two redshift bins and the sample split into two bins of UV luminosity.}
\label{tab:EELG-fraction}
\end{table*}

\subsection{Redshift evolution of the EELG fraction}
\label{sec:method_red_evo}
The [OIII]$\lambda5007$ EW distribution produced in the previous section provides a baseline to probe the redshift evolution of EELGs from z $\sim2$ out into the EOR. Our best fit distribution for z $\sim2$ SFGs produces a mean and a median EW of $168$\,\AA\ and $70$\,\AA\ whereas recent work at z $>7$ show a median EW of $\sim450$\,\AA\ \citep{Labb__2013,deBarros_2019,Endsley20}, suggesting significant evolution in the typical EW in the two billion years between these two epochs. This trend likely reflects evolution in the sSFR and metallicity. In what follows, we seek to use our EW distribution to quantify the redshift evolution in the fraction of extreme emission line galaxies, using the various EW thresholds as discussed in the Section \ref{sec:method_EW_results}.

We will compare our z $\sim2$ measurements to studies of galaxies at higher redshift which also select on rest-frame UV luminosity. We consider three higher redshift studies which characterise rest-optical EW distributions: \citet{Stark2013} at z $\sim4-5$; \citet{Rasappu_2016} at z $\sim5$; and \citet{Endsley20} at z $\sim7$. These three studies are all based on {\it Spitzer}/IRAC colours of Lyman break galaxies, where the flux excess between two adjacent filters is attributed to the presence of strong nebular lines (either [OIII]$\lambda4959,5007$+H$\beta$ or H$\alpha$ depending on the redshift and filter) allowing the rest-optical nebular EWs to be measured for samples of SFGs. Our primary comparison will be to \cite{Endsley20} who model the [OIII]$\lambda4959,5007$ + H$\beta$ EW distribution from 22 Lyman break dropouts at $6.63 < \rm{z} < 6.83$. To compare against EELG fractions from \citet{Endsley20}, we scale the [OIII]$\lambda5007$ EW thresholds into appropriate [OIII]$\lambda4959,5007$ + H$\beta$ thresholds. As described previously, we obtain the [OIII]$\lambda4959,5007$ threshold through a 2.98:1 flux ratio between [OIII]$\lambda5007$ and [OIII]$\lambda4959$ \citep{Storey00}. For H$\beta$, we infer the EW contribution using an H$\beta$:[OIII]$\lambda5007$ EW empirical relation ($\mathrm{log_{10}(H\beta\, EW) = 1.065\,\times\,log_{10}([OIII]\lambda5007\, EW) -0.938}$) obtained from the \citet{Mengtao19} results for a comparable sample of $1.3<\rm{z}<2.4$ EELGs covering a sufficiently broad [OIII]$\lambda5007$ EW range ($\sim100-2500$\,\AA). Here our 500, 750 and 1000\,\AA\, [OIII]$\lambda5007$ EW thresholds are equivalent to 754, 1135 and 1516\,\AA\, [OIII]$+$H$\beta$ EW thresholds.  Our comparison sample also includes two H$\alpha$ EW studies: \cite{Stark2013} at $3.8 < \rm{z} < 5$; and \citet{Rasappu_2016} at $5.1 < \rm{z} < 5.4$. In order to convert our [OIII]$\lambda5007$  EW thresholds into that appropriate for the H$\alpha$ studies, we apply a conversion factor from the linear relation found by \cite{Mengtao19} between H$\alpha$ and [OIII]$\lambda5007$ EW in EELGs. A scaling factor EW(H$\alpha$)/EW([OIII]$\lambda5007$) = 1.0 for [OIII]$\lambda5007$ between 450-800\,\AA\, and 1.1 between 800-2500\,\AA. This mapping suggests that our 500, 750 and 1000\,\AA\, [OIII]$\lambda5007$ EW thresholds are roughly equivalent to 500, 750 and 1100\,\AA\, H$\alpha$ EW thresholds. These conversions are comparable to what has been used in the literature previously (e.g., \citealt{Labb__2013,Rasappu_2016}).

\begin{figure*}
    \centering
    \includegraphics[width=16cm]{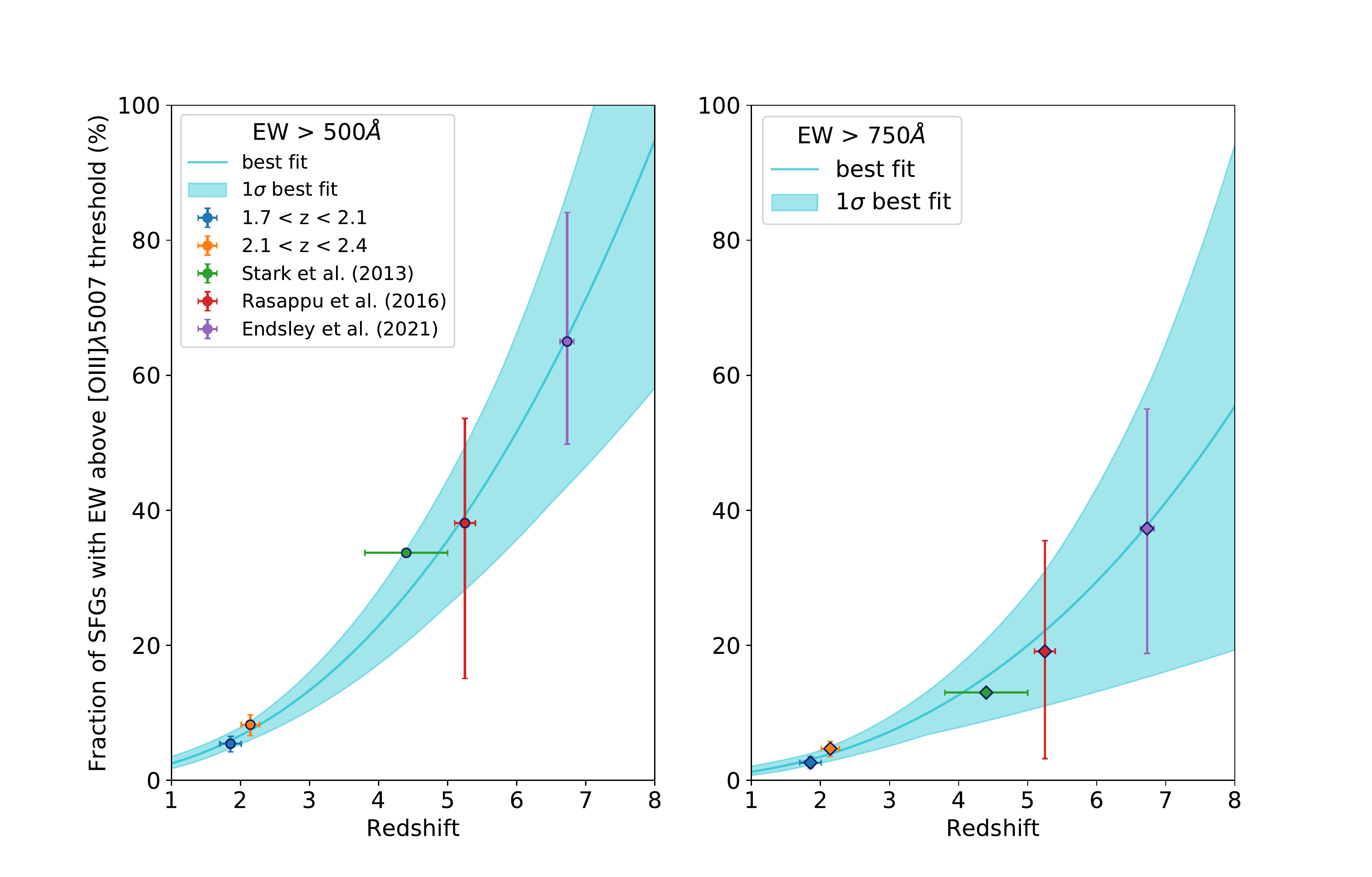}
    \caption{The evolution of the fraction of star-forming galaxies with extremely large [OIII]$\lambda5007$ EW. The fraction for two [OIII]$\lambda5007$ EW thresholds shown in circles (Left plot - 500\,\AA) and diamonds (Right plot - 750\,\AA) across five redshift epochs shown in blue ($1.70 < \rm{z} < 2.01$), orange ($2.01 < \rm{z} < 2.274$), green ($3.8 < \rm{z} < 5.0$), red ($5.1 < \rm{z} < 5.4$) and purple ($6.63 < \rm{z} < 6.83$). Our z $\sim2$ data points are compared to the H$\alpha$ studies at z $\sim5$ and the [OIII]+H$\beta$ study at z $\sim7$ assuming conversions to [OIII]$\lambda5007$ EW. A fitted power law slope $\propto$ (1+z)$^{\rm{P}}$ and associated $1\sigma$ uncertainty are shown in light blue.
    }
    \label{fig:red_ev}
\end{figure*}

To characterise the redshift evolution within our dataset, we divide our z $\sim2$ sample into two redshift bins, $1.70 \leq \rm{z} \leq 2.01$ and $2.01 < \rm{z} \leq 2.274$ containing 329 and 343 galaxies respectively (using spectroscopic redshifts for [OIII] doublet detections and photometric redshifts for non-detections). We report EELG fractions for the two redshift bins in Table \ref{tab:EELG-fraction}. Figure \ref{fig:red_ev} shows the EELG fractions calculated at [OIII]$\lambda5007$ EW thresholds of 500 and 750\,\AA\ and compares to results in the literature at higher redshift.

It is clear from Figure \ref{fig:red_ev} that at higher redshifts a greater proportion of star-forming galaxies are observed in an extreme emission line phase. Not only are the typical SFGs at $\rm{z}\sim7$ ([OIII]$\lambda5007$ EW $\sim500$\,\AA) rare at $\rm{z}\sim2$, representing $5.4^{+1.2}_{-1.1}\%$ and $8.2^{+1.6}_{-1.5}\%$ of the population in our $1.70<\rm{z}<2.01$ and $2.01<\rm{z}<2.274$ bins, but conversely the SFGs typical of $\rm{z}\sim2$ ([OIII]$\lambda5007$ EW $\leq200$\,\AA) are rare at $\rm{z}\sim7$, making up only $\sim3\%$ of SFGs at these high redshifts \citep{Endsley20}. It has been argued that objects with [OIII]$\lambda5007$ EW above 1000\,\AA\ lie above the star-forming main sequence (the relation between the SFR of a galaxy and the stellar mass, see e.g. \citealt{Speagle14}). Such intense line emitters encompass $\sim20\%$ of the population at $\rm{z}>7$, but at $\rm{z}\sim2$ these objects are practically insignificant, with only $1.5^{+0.6}_{-0.4}\%$ ($1.70<\rm{z}<2.01$) and $3.0^{+0.9}_{-0.8}\%$ ($2.01<\rm{z}<2.274$) of SFGs in such a phase in our two intermediate redshfit bins. We consider a power law fit to the redshfit evolution of the EELG fraction (frac(z)$\,=\,$frac$_{0}$\,(1+z)$^{\rm{P}}$) and find $\rm{z}=0$ fraction and power law slope parameters (frac$_{0}$=0.47$^{+0.12}_{-0.10}\%$, P=2.42$^{+0.18}_{-0.18}$) for an EW threshold of 500\,\AA\ and (frac$_{0}$=0.22$^{+0.11}_{-0.07}\%$, P=2.51$^{+0.27}_{-0.32}$) for an EW thresholds of 750\,\AA. 

\begin{table}
\begin{center}
\caption{Best fit model parameters for the rest-frame EW(\,\AA). We show the results for the log-normal distribution (LN) and normal distribution in logarithmic space (N) for our two redshift sub-samples.}
\label{tab:mcmc_results3}
\begin{tabular}{cccccc}
z-range  & N  & $\mu_{\rm{LN}}$        & $\sigma_{\rm{LN}}$    & $\mu_{\rm{N}}$ & $\sigma_{\rm{N}}$ \\ \hline
\hline
$1.70 - 2.01$ & 329 & $4.26^{+0.09}_{-0.09}$ & $1.21^{+0.08}_{-0.07}$ & $1.21^{+0.12}_{-0.11}$ & $0.53^{+0.03}_{-0.03}$\\
\\
$2.01 - 2.274$ & 343 & $4.26^{+0.11}_{-0.12}$ & $1.43^{+0.11}_{-0.10}$ & $0.97^{+0.18}_{-0.18}$ & $0.62^{+0.05}_{-0.04}$\\
\end{tabular}
\end{center}
\end{table}

\subsection{The dependence of the EELG fraction on UV luminosity}
\label{sec:method_uv_lum} 
Here we characterise the luminosity-dependence of the [OIII]$\lambda5007$ EW distribution in the range sampled by our dataset ($-21.6 < \rm{M_{\rm{UV}}} < -19$). At z $\simeq 7$, an analysis using {\it Spitzer}/IRAC flux excesses as a probe of rest-optical line strengths found no evidence for a significant [OIII]+H$\beta$ EW trend with M$_{\rm{UV}}$ \citep{Endsley20}. Given the close connection between [OIII]$\lambda5007$ EW and the ionizing efficiency (see discussion in Section \ref{sec:intro}), this result has implications for the M$_{\rm{UV}}$-dependent contribution of galaxies to reionisation.  Physically we may expect the [OIII]$\lambda5007$ EW to be stronger towards lower UV luminosities given the correlation between M$_{\rm{UV}}$ and stellar mass at z $\simeq 2$ (e.g., \citealt{Reddy09}) and the relationship between mass and metallicity (e.g., \citealt{Sanders21}). The larger electron temperature in lower metallicity systems can act to boost collisionally-excited emission lines such as [OIII]$\lambda5007$ \citep{Reddy18}. The [OIII]$\lambda5007$ EW additionally depends  on the sSFR \citep[e.g.,][]{Mengtao19}. This is especially true at very large sSFR, where the  [OIII]$\lambda5007$ EW is enhanced by the weak underlying rest-optical continuum associated with very young stellar populations. If the large sSFR phase  is more common in galaxies with lower UV luminosities (as might be expected if bursts are more common in lower luminosity and lower mass systems), we would expect to see larger EELG fractions at the faint end of the luminosity function.

To investigate whether the [OIII]$\lambda5007$ EW distribution varies with UV luminosity, we first separate our sample into a UV-bright ($-21.6 < \rm{M_{\rm{UV}}} < -19.5$, 337 galaxies) and UV-faint sub-sample ($-19.5<\rm{M_{\rm{UV}}}<-19$, 335 galaxies). We recompute the best-fit log-normal parameters for the [OIII]$\lambda5007$ EW distributions in both magnitude bins. The results reveal broad consistency in the model mean and also the median of the two $\rm{M_{\rm{UV}}}$ bins: the mean EW values of the UV-faint and UV-bright samples are $165.1^{+2.0}_{-1.5}$ and $174^{+1.4}_{-1.1}$\,\AA\ respectively (where we quote the standard error on the mean), and the median values are $63^{+8}_{-7}$ and $74^{+7}_{-6}$\,\AA. The full set of best-fit parameters for both sub-samples is presented in Table \ref{tab:mcmc_results4}. As can be seen in the Table, both the average and width of the EW distributions are consistent within $2\sigma$.

\begin{figure*}
    \centering
    \includegraphics[width = 15cm]{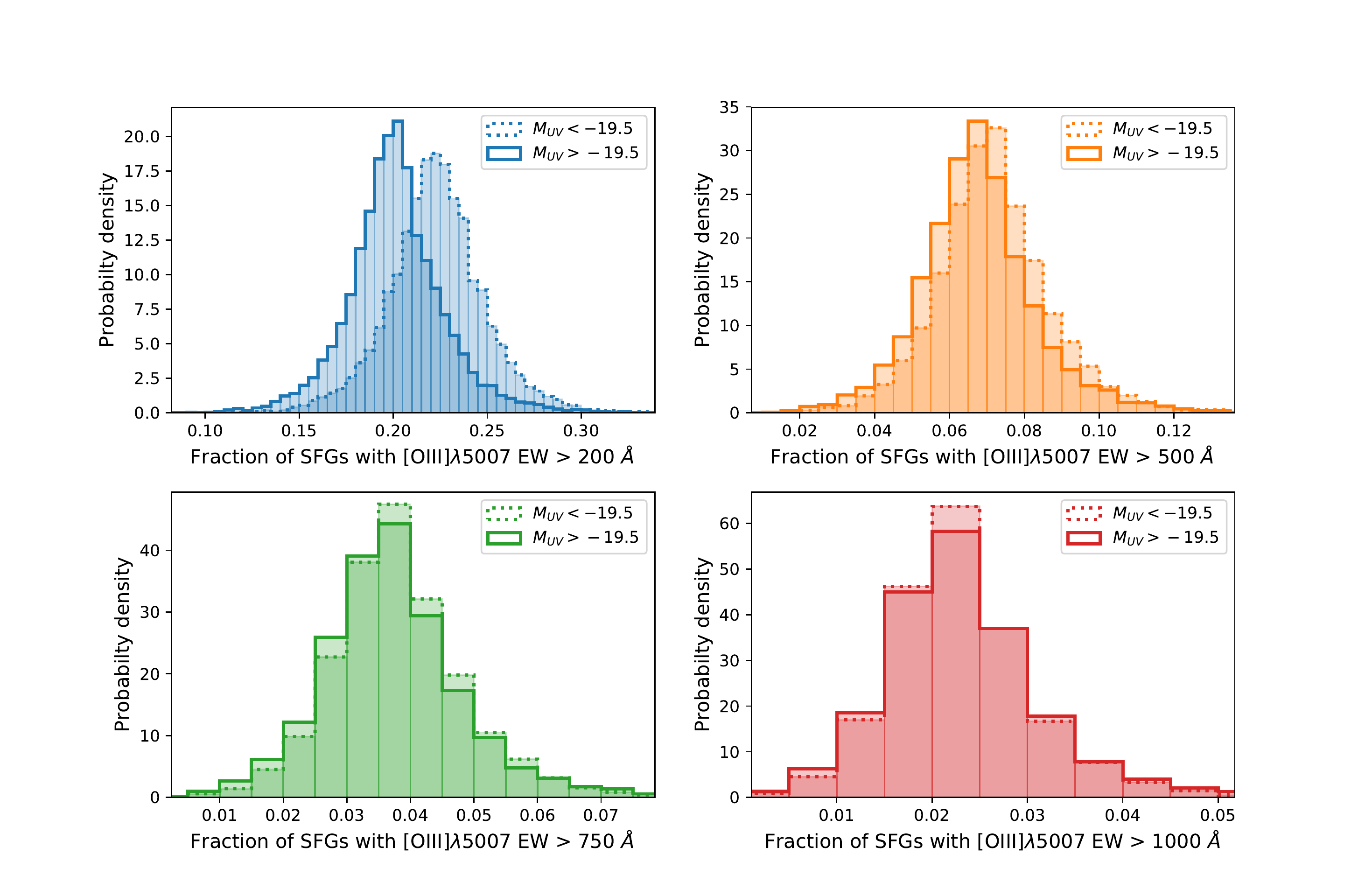}
    \caption{The fraction of $M_{\rm{UV}}$ bright and $M_{\rm{UV}}$ faint star-forming galaxies at z $\sim2$ with extremely large [OIII]$\lambda5007$ EW. Posterior distributions for the bright $-21.6 < M_{\rm{UV}} < -19.5$ (solid lines) and faint $-19.5 < M_{\rm{UV}} < -19.0$ (dashed lines) shown for four different [OIII]$\lambda5007$ EW thresholds in blue (200\,\AA), orange (500\,\AA), green (750\,\AA) and red (1000\,\AA). }
    \label{fig:Muv_evolution}
\end{figure*}

We also consider the luminosity-dependence of the EELG population, computing the fraction of galaxies that have [OIII]$\lambda5007$ EWs above four physically motivated thresholds (200, 500, 750 and 1000\,\AA\,, see Table \ref{tab:EELG-fraction}). The EELG fraction posterior density functions for both UV bins are shown in Figure \ref{fig:Muv_evolution}.  For each threshold (each sub-panel in Figure \ref{fig:Muv_evolution}) the derived fraction of SFGs with an [OIII]$\lambda5007$ EW above the given threshold is consistent between the two luminosity sub-samples. The fraction of galaxies with an EW above the typical [OIII]$\lambda5007$ EW at z $\simeq 2$ (200\,\AA) is similar in the fainter and brighter bins (from $20.1^{+2.4}_{-2.1}\%$ to $22.3^{+2.4}_{-2.3}\%$), and the same is found for the relative abundance of the most extreme emission line galaxies (EW $\geq1000$\,\AA) ($2.3^{+0.8}_{-0.7}\%$ and $2.2^{+0.7}_{-0.6}\%$ for the fainter and brighter bins). The absence of significant variations in the [OIII]$\lambda5007$ EW distributions between the UV-bright and UV-faint bins is consistent with the findings of \citet{Endsley20} at z $\simeq 7$. A larger dynamic range in M$_{\rm{UV}}$ may be required to find a trend between [OIII]$\lambda5007$ EW and UV luminosity.

\begin{table}
\begin{center}
\caption{Best fit model parameters for the rest-frame EW(\,\AA). We show the results for the log-normal distribution (LN) and normal distribution in logarithmic space (N) for our two  $M_{\rm{UV}}$ sub-samples.}
\label{tab:mcmc_results4}
\begin{tabular}{ccccccc}
$M_{\rm{UV}}$  & N  & $\mu_{\rm{LN}}$        & $\sigma_{\rm{LN}}$    & $\mu_{\rm{N}}$ & $\sigma_{\rm{N}}$ \\ \hline
\hline
$-19.5$ to $-19.0$ & 335 & $4.13^{+0.12}_{-0.12}$ & $1.39^{+0.12}_{-0.11}$ & $0.95^{+0.19}_{-0.18}$ & $0.60^{+0.05}_{-0.05}$\\
\\
$-21.6$ to $-19.5$ & 337 & $4.31^{+0.08}_{-0.09}$ & $1.30^{+0.08}_{-0.07}$ & $1.14^{+0.13}_{-0.12}$ & $0.57^{+0.03}_{-0.03}$
\end{tabular}
\end{center}
\end{table}

\section{Discussion}
\label{sec:discussion}
In this paper, we have derived the [OIII]$\lambda$5007 EW distribution in z $\simeq 2$ UV-selected galaxies, thereby quantifying the fraction of EELGs at $\rm{z}\simeq 2$. Locally (z $\sim0$) and at intermediate redshifts ($1<\rm{z}<3$), EELGs have been shown to be extremely efficient ionisers, both due to their large ionisation production efficiencies \citep{Chevallard_2018, Mengtao19, Nakajima19, Emami20} and their large escape fractions \citep[e.g.][]{Jaskot19, Nakajima19, Izotov16, Vanzella16,Fletcher19}. However the rareness of this population suggests they are likely to make a sub-dominant contribution to the ionising background from star-forming galaxies at $\rm{z}\simeq 2$. In this section, we estimate the fractional contribution from EELGs at $\rm{z}\simeq 2$ to the ionising background, combining [OIII]$\lambda5007$ EW distributions with nominal assumptions on the ionising efficiency of the population.

We first consider the ionising contribution of EELGs at $\rm{z}\simeq 2$, focusing on those systems with [OIII]$\lambda5007$ EW $>$ 750\,\AA. While this subset comprises just 3.6\% of the $\rm{z}\simeq 2$ population (Table \ref{tab:EELG-fraction}), they are thought to be very efficient ionisers.  These objects have typical ionising photon production efficiencies of log$_{10}(\xi_{\rm{ion}},$ erg s$^{-1}$Mpc$^{-3}$Hz$^{-1}) = 25.58$ \citep{Mengtao19}, 3.3$\times$ greater than that in more typical star-forming galaxies at $\rm{z}\simeq 2$ \citep{Shivaei_2018}. While not all EELGs show ionizing photon leakage, values are often significant in the population, with systems having the largest [OIII]$\lambda$5007 EWs (i.e., $>$750\,\AA) often found with estimated escape fractions of up to 20-50\% \citep{Vanzella16,RiveraThorsen17,Izotov18,Fletcher19}. For the purposes of this calculation, we will assume that this range of $20-50\%$ escape fractions is exhibited by half of galaxies with [OIII]$\lambda5007$ EW $>$ 750\,\AA, with the remaining half leaking no ionising radiation. While this f$_{\rm{esc}}$ distribution is clearly still very uncertain, the values are broadly consistent with known constraints on LyC escape fractions and indirect indicators of leakage in this extreme [OIII] emitting population (e.g., \citealt{Izotov18,Jaskot19,Tang20,Du20}). In what follows, we take these values and calculate the comoving emissivity of ionising photons ($\dot{n}_{\rm{ion}}$, s$^{-1}$Mpc$^{-3}$) for these intense EELGs, but we caution that improved distributions of escape fractions in this population are ultimately required for more confident inferences. We focus on galaxies in the luminosity range probed in this paper ($-21.6 < M_{\rm{UV}} < -19$) and assume the \citet{Reddy09} UV luminosity function. We multiply the far UV luminosity density by the fraction of galaxies with [OIII]$\lambda5007$ EW $>$ 750\,\AA\ (3.6\%). After accounting for the ionising production efficiency and the range of escape fractions, we find that this population injects a comoving ionising photon emissivity of between 3 and 8$\times10^{49}$ s$^{-1}$ Mpc$^{-3}$ into the IGM at $\rm{z}\simeq 2$ for our assumed parameters. We can now estimate the fractional contribution these EELGs make to the total ionising background produced by $\rm{z}\simeq 2$ star-forming galaxies. To do so, we use population-average estimates of the ionising photon production efficiency and escape fraction. For the escape fraction, we use the recently-derived value from \citet{Pahl21}, indicating an average of $f_{\rm{esc}} \sim6\%$ for $\rm{z}\simeq 3$ UV-selected galaxies (see also \citealt{steidel18, Bassett21}).  Here we assume that $\rm{z}\simeq 2$ galaxies have a similar value.  For the ionising production efficiency, we use the value derived for the $\rm{z}\simeq 2$ UV-selected population, taking the same attenuation law as we assumed for EELGs \citep{Shivaei_2018}. The estimated ionising emissivity from UV-selected galaxies over our luminosity range is then $\dot{n}_{\rm{ion}}=1.5\times10^{50}$ s$^{-1}$Mpc$^{-3}$. While the emissivities quoted above are nominal estimates with significant uncertainties, they illustrate that EELGs do indeed make a sub-dominant contribution to ionising output from galaxies at $\rm{z}\simeq 2$. However if the large escape fractions assumed here for a subset of the most intense of EELGs are correct, it would indicate that even at $\rm{z}\simeq 2$, this population makes a non-negligible contribution to the ionising background of star-forming galaxies. 

The ionising contribution from EELGs is likely to increase substantially as we enter the reionisation era. Given current constraints on the $\rm{z}\simeq 7$ [OIII]+H$\beta$ EW distribution from \citet{Endsley20} and nominal assumptions about the relation between [OIII]+H$\beta$ and [OIII]$\lambda5007$ EWs (see Section \ref{sec:method_red_evo}), the percentage of star-forming galaxies with [OIII]$\lambda5007$ EW >750\,\AA\ increases by a factor of 10 between $\rm{z}\simeq 2$ and $\rm{z}\simeq 7$, from 3.6\% ($\rm{z}\simeq 2$) to close to 37\% ($\rm{z}\simeq7$). The percentage of those with an [OIII]$\lambda$5007 EW >500\,\AA\ also increases by a factor of 10 between $\rm{z}\sim2-7$, reaching close to 65\% of the population at $\rm{z}\simeq 7$ \citep{Endsley20}. So while the majority of the $\rm{z}\sim2$ star-forming galaxy ionising background comes from more typical modes of star formation than EELGs, during the reionisation era the bulk of the ionising photons will likely come from EELGs. Future work is required to test if the ionising efficiency of this population evolves with redshift. The first spectra of EELGs in the reionisation era suggest similarly intense radiation fields as are often seen at lower redshifts \citep[e.g.,][]{Stark15a, Stark15b, Stark17, Mainali17, Schmidt17, Hutchison19, Jiang21, Topping21}, but {\it JWST} will soon allow much-improved investigation of the ionising output of reionisation-era EELGs.

The rapid evolution in the EELG population is suggestive of a 
shift in the main star forming mode between $\rm{z}\sim2$ and $\rm{z}\sim7$. The most intense EELGs are likely in the midst of a burst or recent upturn in star formation. The increase in this population with redshift may suggest that such intense bursts are becoming more common in the reionisation era. This is perhaps consistent with the observed rise in the galaxy merger rate and specific mass accretion rate observed between $1 < \rm{z} < 6$ \citep{Duncan19}, both of which may spark and feed more frequent and stronger bursts of star formation. The nebular rest-optical line EW distributions (and the related sSFR distributions) encode useful information on the star formation history, with the tails of the distribution (both at high and low sSFR) constraining the strength and duty cycle of bursts. The evolution and mass-dependence of these distributions will soon be constrained in more detail by {\it JWST}. Direct comparison of these observations to simulations and semi-analytic models promises valuable insight into the presence of bursts in the earliest galaxies.

\section{Summary}
Recent years have seen increased interest in extreme emission line galaxies, owing to their efficiency as ionising agents and their apparent ubiquity in the reionisation era. The [OIII]$\lambda5007$ equivalent width distribution constrains the percentage of star-forming galaxies at a given epoch caught in an extreme emission line phase. While efforts have begun to characterise the distribution of [OIII]+H$\beta$ line strengths at $\rm{z}\simeq 7$ \citep{Labb__2013,Smit_2014,deBarros_2019,Endsley20}, similar measurements do not exist at $\rm{z}\simeq 2$, impeding efforts to track the redshift evolution of the EELG population. We establish the best fit log-normal model for the [OIII]$\lambda5007$ equivalent width distribution in a rest-UV selected sample ($M_{\rm{UV}} < -19$) in the redshift range $1.700 < \rm{z} < 2.274$, using the combination of HDUV photometry and the 3D-HST grism spectra. With the [OIII]$\lambda$5007 EW distribution, we quantify the fraction of $\rm{z}\simeq 2$ galaxies with extreme line emission, providing the low redshift baseline necessary to characterise the evolution of this population. The fraction of UV-selected galaxies with an [OIII]$\lambda5007$ EW above  $200, \,500, \,750$ and 1000\,\AA\ is found to be $21.2^{+1.7}_{-1.6}\%, 6.8^{+1.0}_{-0.9}\%, 3.76^{+0.7}_{-0.6}\%$ and $2.2^{+0.5}_{-0.4}\%$ respectively. We find no strong evidence that the EELG fractions vary with UV luminosity in the range considered in this paper ($-21.6 < M_{\rm{UV}} < -19.0$), consistent with results at $\rm{z}\simeq 7$ \citep{Endsley20}. 

Comparison to results at higher redshift (e.g., \citealt{deBarros_2019,Endsley20}) reveals rapid redshift evolution, with the fraction of galaxies having [OIII]$\lambda$5007 EW$>500$\,\AA\ increasing from 6.8\% at $\rm{z}\simeq 2$ to 65$\%$ at $\rm{z}\simeq 7$ (for nominal assumptions about the H$\beta$ contribution at $\rm{z}\simeq 7$).  We find a similar increase with a slightly higher [OIII]$\lambda$5007 EW  threshold ($>750$\,\AA), with 3.6\% of the population in this regime at $\rm{z}\simeq 2$ and 37\% at $\rm{z}\simeq 7$. Even accounting for their enhanced ionising efficiency, EELGs are too rare at z $\sim2$ to dominate the ionising background produced by star-forming galaxies. However, a far greater percentage of galaxies will be in an extreme emission line phase at $\rm{z}\simeq 7$, providing an ideal population for ionising the intergalactic medium. Future work will soon offer much-improved measures of the evolving EELG population, both in terms of their ionising efficiency and their mass-dependent contribution to the total galaxy population. These studies promise valuable insights into the contribution of galaxies to reionisation and the redshift and mass-dependence of bursts in early galaxies.  

\section*{Acknowledgements}
KB acknowledges funding from the Science and Technology Facilities Council (STFC) Grant Code ST/R505006/1. AJB has received funding from the European Research Council (ERC) under the European Union’s Horizon 2020 Advanced Grant 789056 \lq\lq First Galaxies". This work is based on observations taken by the 3D-HST Treasury Program (HST-GO-12177 and HST-GO-12328) with the NASA/ESA Hubble Space Telescope, which is operated by the Association of Universities for Research in Astronomy, Inc., under NASA contract NAS5-26555.

\section*{Data Availability}
This work is based on public data available from the HST archive, and we have used the data products released by 3D-HST and HDUV. The catalogues and analysis routines used in this work are available upon reasonable request to the authors.




\bibliographystyle{mnras}
\bibliography{reference} 




\bsp	
\label{lastpage}
\end{document}